\documentclass[preprint,review,12pt]{elsarticle}
\usepackage{amssymb}
\usepackage{graphicx}% Include figure files
\usepackage{dcolumn}% Align table columns on decimal point
\usepackage{bm}% bold math
\usepackage{pifont}
\usepackage{natbib}
\usepackage{threeparttable}

\journal{Journal of Solid State Chemistry}

\begin{document}

\begin{frontmatter}

\title{Structural and physical properties of layered oxy-arsenides \textit{Ln}RuAsO (\textit{Ln} = La, Nd, Sm, Gd)}

\author{Michael A. McGuire}
\cortext[MAM]{Corresponding author. \textit{E-mail address:} McGuireMA@ornl.gov (M. A. McGuire)}
\author{Andrew F. May \corref{}}
\author{Brian C. Sales}

\address{Oak Ridge National Laboratory, Oak Ridge, Tennessee 37831 USA}

\begin{abstract}
Polycrystalline samples of LaRuAsO, NdRuAsO, SmRuAsO, and GdRuAsO have been synthesized and studied using powder x-ray diffraction, electrical transport, magnetization, and heat capacity measurements. Variations in structural properties across the series reveal a trend toward more ideal tetrahedral coordination around Ru as the size of the rare earth element is reduced. The lattice parameters of these Ru compounds show a more anisotropic response to variation in $Ln$ than their Fe analogues, and significant anisotropy in thermal expansion is also observed. Transport measurements show metallic behavior, and carrier concentrations near 10$^{21}$--10$^{22}$ electrons per cm$^3$ are inferred from simple analysis of Hall effect measurements. Anomalies in resistivity, magnetization, and heat capacity indicate antiferromagnetic ordering of rare earth moments at 5 K for GdRuAsO, 4.5 K for SmRuAsO, and $<$2 K for NdRuAsO. Magnetization measurements on LaRuAsO show no evidence of a magnetic moment on Ru. Observed behaviors are compared to those reported for similar Fe and Ru compounds.
\end{abstract}

%\begin{keyword}
%LaRuAsO \sep NdRuAsO \sep SmRuAsO \sep GdRuAsO \sep crystal structure \sep heat capacity \sep transport properties \sep magnetization \sep

%\end{keyword}

\end{frontmatter}

\newpage

\section{Introduction}

Rare earth transition metal oxy-arsenides adopting the ZrCuSiAs structure-type (1111 materials) contain square nets of transition metal atoms coordinated by arsenic in an edge-sharing tetrahedral geometry \cite{Quebe}. These materials, and others containing the same transition metal--arsenic layers, have been a main focus of experimental and theoretical studies in the solid state chemistry and condensed matter physics communities since the discovery of high temperature superconductivity in many materials, primarily containing iron, and several structural families \cite{Kamihara, Johnston-review, Sefat-review}. A common theme, and perhaps necessary condition, for high temperature superconductivity in the iron compounds is the suppression of magnetism, which is accomplished by chemical substitutions or application of pressure. While aleovalent substitution (doping) is most common, replacing some Fe with isovalent Ru has been shown to produce superconductivity in the family of layered iron arsenides adopting the ThCr$_2$Si$_2$ structure type (122 materials) \cite{Schnelle-SrFeRuAs, Sharma-BaFeRuAs}. However, this substitution does not produce superconductivity in 1111 materials, although it does suppress the magnetism \cite{McGuire-PrFeRuAsO, Pallecchi-LaFeRuAsO, Pal-SmGdFeRuAsO}. Indeed, partial replacement of Fe with Ru in already superconducting compositions (SmFeAsO$_{1-x}$F$_{x}$) decreases the superconducting critical temperature \cite{Tropeano-SmFeRuAsOF}. The response to Ru substitution is one of the few striking differences between the behavior of 1111 and 122 materials.

The natural extension of these substitution studies is the analysis of the pure ruthenium compounds. The 122 materials SrRu$_2$As$_2$ and BaRu$_2$As$_2$ have been studied and are diamagnetic metals \cite{Jeitschko-LaRu2P2, Nath-AeRu2As2, Schnelle-SrFeRuAs}. Interestingly, the 122 phosphide LaRu$_2$P$_2$, with trivalent lanthanum in place of the ususal divalent alkaline earth, is superconducting below 4.1 K \cite{Jeitschko-LaRu2P2}. Literature reports for $Ln$RuAsO ($Ln$ = lanthanide) are limited to lattice constants \cite{Quebe} and resistivity for $Ln$ = La and Ce \cite{Chen-LnRuAsO}. A more thorough investigation of the Ru materials is important in developing a full understanding of the behavior of these interesting chemical systems.

The current study aims to examine the evolution of structural and basic physical properties of some 1111 materials of composition $Ln$RuAsO as the lanthanide ($Ln$) is varied. Full crystal structure refinements at room temperature and thermal expansion, heat capacity, electrical transport, and magnetic behavior below 300 K are reported for LaRuAsO, NdRuAsO, SmRuAsO, and GdRuAsO. The results are compared to similar 1111 and 122 materials containing Fe and Ru.

\section{Experimental Details}
RuAs was made by reacting reduced Ru powder with As pieces in an evacuated silica ampoule at 1000 $^\circ$C and used as a starting material for the target compounds. Polycrystalline samples of LaRuAsO, NdRuAsO, SmRuAsO, and GdRuAsO were synthesized from thoroughly ground mixtures of RuAs with fresh $Ln$ filings and dry $Ln_2$O$_3$ powder. The starting materials were handled and mixed inside a He glove box. Samples ($\sim$ 2 g each) were pressed into 1/2 inch diameter pellets and placed in covered alumina crucibles inside silica tubes. The tubes were evaculated, back-filled with $\sim$0.2 atm of ultra-high-purity Ar, and flame sealed. The samples were heated at 1200--1250 $^\circ$C for 12--36 hours several times, and were ground and pelletized between the heating cycles. Surface contamination from reaction with vapor from the SiO$_2$ tubes was removed at each step.

Powder x-ray diffraction (PANalytical X'Pert Pro MPD, monochromatic Cu-K$\alpha_1$ radiation) was used to determine phase purity and refine the crystal structures using the program Fullprof \cite{Fullprof}. Low temperature powder diffraction was performed with an Oxford Phenix closed-cycle cryostat. A Quantum Design Physical Property Measurement System was used for transport and heat capacity measurements, and a Quantum Design Magnetic Property Measurement System SQUID magnetometer was used for magnetization measurements.

\begin{figure}
\includegraphics[width=3.25in]{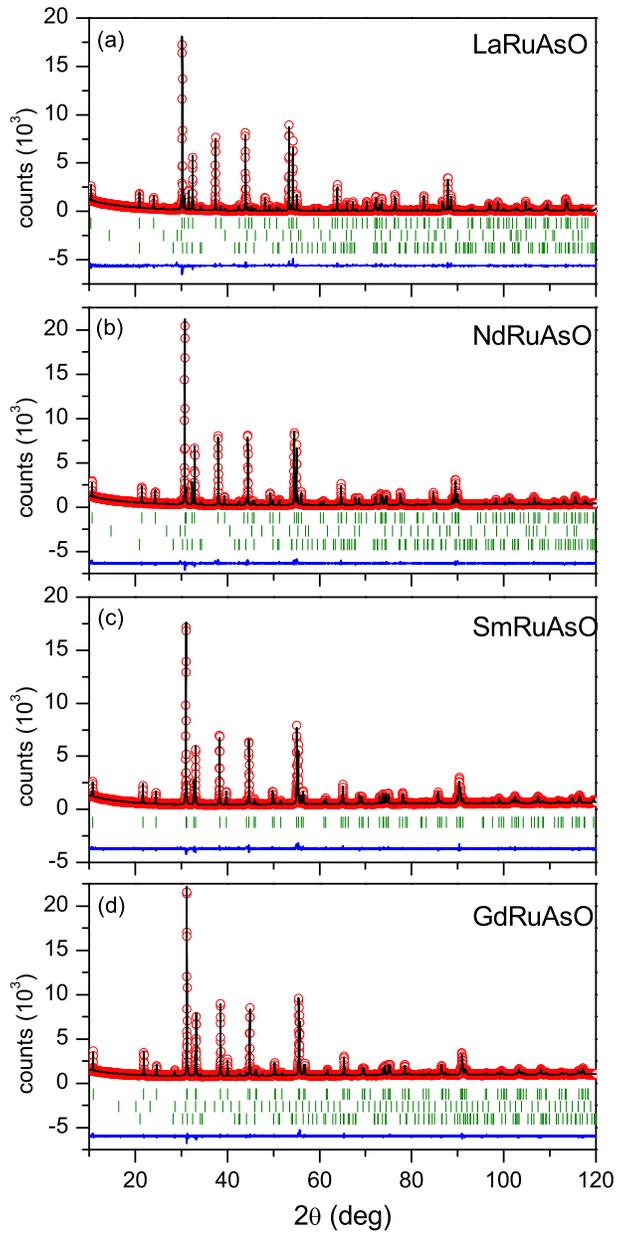}
\caption{\label{fig:rietveld}
Reitveld refinement of room temperature powder x-ray diffraction data. Measured intensities are shown as circles, calculated intensities as grey lines, and difference curves as black lines. The upper set of ticks in each panel correspond to the main phase. Middle and lower ticks in (a,b,d) represent $Ln_2$O$_3$ and RuAs, respectively.
}
\end{figure}

Room temperature Rietveld refinements are shown in Figure \ref{fig:rietveld}, and indicate all samples are $\gtrsim$ 90 \% pure. Impurities including $Ln_2$O$_3$ and RuAs were observed in the samples with $Ln$ = La, Nd, and Gd; however, no significant impurity peaks were observed in the SmRuAsO sample. Agreement factors for the fits ranged from $R_p$ = 2.9--7.1, $R_{wp}$ = 3.7--10.1. $\chi^2$ = 1.5--2.2.

\section{Results and Discussion}

\subsection{Structural properties}
\begin{figure}
\includegraphics[width=3.25in]{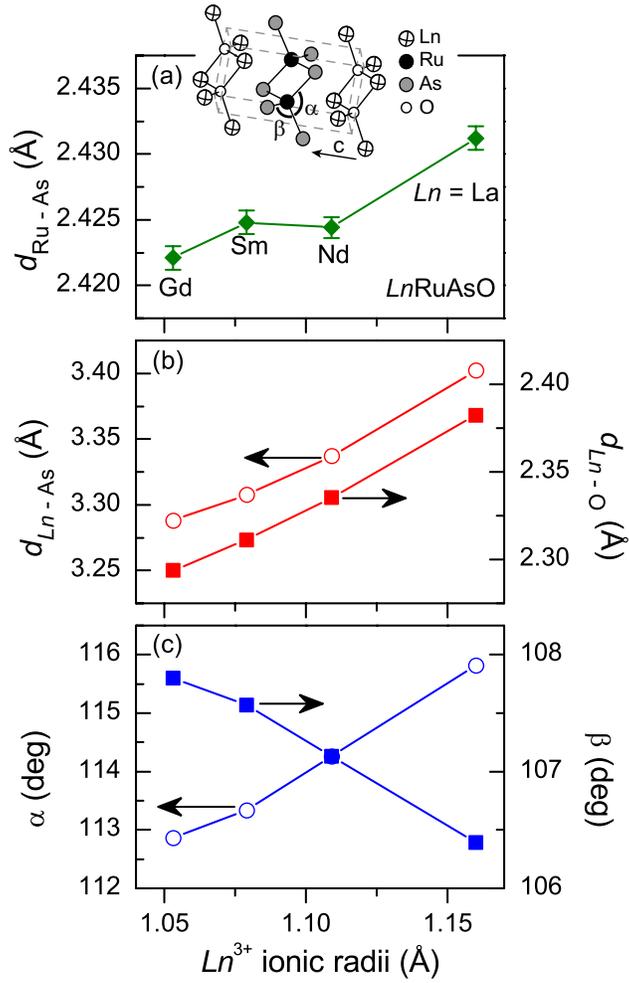}
\caption{\label{fig:bonds}
Structural parameters of \textit{Ln}RuAsO plotted as a function of the ionic radius of $Ln^{3+}$ in eight-fold coordination \cite{Shannon}. (a) Ru--As distance, with the structure shown in the inset. (b) $Ln$--As and $Ln$--O distances. (c) Coordination angles around Ru as defined in the inset of (a). The two angles are not independent due to the Ru site symmetry (-4\textit{m}2).
}
\end{figure}

These materials adopt the ZrCuSiAs structure type shown in Figure \ref{fig:bonds}a, with $Ln$ and As at Wyckoff positions 2c ($\frac{1}{4}$ $\frac{1}{4}$ $z$), Ru at 2a ($\frac{3}{4}$ $\frac{1}{4}$ 0), and O at 2b ($\frac{3}{4}$ $\frac{1}{4}$ $\frac{1}{2}$). Room temperature structural parameters are listed in Table \ref{lattice-table}. The lattice constants are in good agrement with the original report of these materials, and reflect the expected lanthanide contraction, as previously noted \cite{Quebe}. Table \ref{lattice-table} also lists refined atomic positions $z_{As}$ and $z_{Ln}$, which have not been previously reported.

\begin{table}
\begin{center}
\caption{\label{lattice-table} Room temperature lattice constants and z-coordinates for the $Ln$ and As positions from powder x-ray diffraction. Uncertainties on the last digit of refined parameters are listed in parentheses, and are those reported by the refinement program (Fullprof). Hall coefficient ($R_H$) measured at 2 K. Effective moment ($\mu_{eff}$), Weiss temperature ($\theta$), and Neel temperature (T$_N$) for compounds with magnetic rare earth elements determined from temperature dependent magnetization and heat capacity measurements.}
\begin{tabular}{lcccc}
\hline									
	&	LaRuAsO	&	NdRuAsO	&	SmRuAsO	&	GdRuAsO	\\
\hline									
$a (\AA)$	&	4.11954(2)	&	4.07259(2)	&	4.05194(3)	&	4.03632(3)	\\
$c(\AA)$	&	8.49128(5)	&	8.28808(6)	&	8.19367(8)	&	8.12406(7)	\\
$z_{Ln}$	&	0.1409(1)	&	0.1379(1)	&	0.1358(1)	&	0.1342(1)	\\
$z_{As}$	&	0.6521(2)	&	0.6588(2)	&	0.6626(2)	&	0.6649(2)	\\

$\mu_{eff}$ ($\mu_B$)	&	--	&	3.5	&	--	&	8.2	\\
$\theta$ (K)	&	--	&	-14	&	--	&	-20	\\
$T_N$ (K)	&	--	&	$<$ 2	&	4.5	&	5.0	\\

$R_{H}^{\,\,2K}$ &	$-3.1\times10^{-3}$	&	$-1.9\times10^{-3}$	&	$-8.2\times10^{-4}$	&	 $-3.2\times10^{-4}$ \\
($cm^{-3}/C$) 	\\
\hline									
\end{tabular}
\end{center}
\end{table}

Based on the full structure refinements, the evolution of interatomic distances and angles can be examined as $Ln$ is varied. Selected distances and angles are plotted in Figure \ref{fig:bonds}. As expected, the Ru--As distance is least sensitive to the identity of $Ln$ (Figure \ref{fig:bonds}a), changing by less than 0.4 \% across this series. The $Ln$--As and $Ln$--O distances (Figure \ref{fig:bonds}a) increase smoothly as the $Ln$ ionic radius increases. Figure \ref{fig:bonds}c shows the two As--Ru--As bond angles, as defined in the inset of Figure \ref{fig:bonds}a. The observation that $\alpha\ > \beta$ indicates that the As tetrahedron around Ru is compressed along the c-axis. As the $Ln$ radii is decreased along the series from La to Gd, $\alpha$ decreases and $\beta$ increases, moving toward ideal tetrahedral coordination. It is interesting to note that a strong correlation between these angles and superconducting critical temperatures has been observed in the related Fe compounds, where the highest critical temperatures occur when the tetrahedra are closest to ideal \cite{Zhao-Ce, Kreyssig}.

A systematic structural study of the analogous Fe compounds using single crystal x-ray diffraction results has been reported \cite{Nitsche-LnFeAsO}. Comparison with Table \ref{lattice-table} and Figure \ref{fig:bonds} shows that the same structural trends exist in $Ln$FeAsO and $Ln$RuAsO. In the Fe compounds, all the interatomic distances are shorter than those in the corresponding Ru analogues. $Ln$--As and $Ln$--O distances in the Fe compounds are shorter by about 0.3--0.4 and 0.2 $\AA$ for the lanthanides studied here. This reflects the compression of the overall structure when Ru is replaced by smaller Fe, and is often referred to as chemical pressure. The Fe--As distances in $Ln$FeAsO vary from 2.412 to 2.392 $\AA$ from $Ln$ = La to Gd, showing a much stronger dependence on $Ln$ than the Ru--As distances in Figure \ref{fig:bonds} which vary by less than 0.01 $\AA$ across the same series. These results suggest that the Ru compounds are in some sense less compressible than the Fe analogues, at least in the $ab$-plane. This may indicate the RuAs layers are stiffer than the FeAs layers due to increased Ru--Ru repulsion between the larger Ru atoms. This is consistent with the changes in lattice constants across the $Ln$ series. From LaRuAsO to GdRuAsO $a$ changes by 2.0 \% and $c$ by 4.3 \%. From LaFeAsO to GdFeAsO $a$ changes by 2.9 \% and $c$ by 3.2 \%. The $Ln$RuAsO lattice response is much more anisotropic than that of $Ln$FeAsO, and it is stiffer in the $a$ direction than in $c$. In addition, the tetrahedral coordination around Ru (Figure \ref{fig:bonds}c) is significantly more flattened along $c$ than that found around Fe in $Ln$FeAsO \cite{Nitsche-LnFeAsO}, also pointing to Ru-Ru repulsion in the $ab$-plane. Finally, it is interesting to note that the deviation from a linear trend in Ru-As distance at $Ln$ = Nd in $Ln$RuAsO (Figure \ref{fig:bonds}a) is also seen in $Ln$FeAsO \cite{Nitsche-LnFeAsO}, although its origin and significance remain unclear.

\begin{figure}
\includegraphics[width=3.0in]{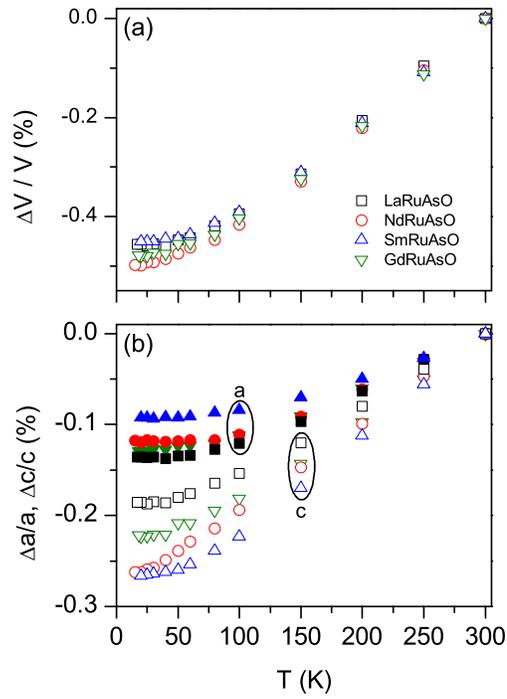}
\caption{\label{fig:thermal-expansion}
Temperature dependence of (a) the unit cell volume (V) and (b) the lattice constants $a$ (solid symbols) and $c$ (open symbols) determined by powder x-ray diffraction and normalized to values at 300 K. Uncertainties based on the Rietveld fitting are smaller than the data markers.
}
\end{figure}

The temperature dependence of the lattice parameters and unit cell volume between $\sim$20 and 300 K are presented in Figure \ref{fig:thermal-expansion}. The data are normalized to 300 K values. All four compounds show similar volume thermal expansion behavior. A small divergence in $\Delta$V/V among materials occurs at the lowest temperatures, but no simple relationship to the composition is observed. The temperature dependence of the lattice constants shows strong anisotropy in all materials, with $c$ changing more rapidly than $a$. This reflects the layered nature of the crystal structure which contains covalently bonded RuAs layer extending in the $ab$-plane, and is consistent with the trends in lattice constant with $Ln$ at room temperature discussed above. The magnitude of this anisotropy ([$\Delta c/c$] / [$\Delta a/a$]) depends strongly on the composition, and reaches a value of 1.4, 2.2, 2.9, and 1.7 at 20 K for La, Nd, Sm, and Gd, respectively. The reason for the maximum for SmRuAsO is not clear. However, among the lanthanides studied here Sm does possess two unique properties: (1) a tendency toward mixed-valent behavior (Sm$^{2+/3+}$), (2) two low energy, closely spaced magnetic configurations in the trivalent state (J = 5/2, 7/2). These can be expected to affect the temperature dependence of physical and structural properties. It is interesting to speculate whether this increased ``flexibility'' of Sm could be the reason that the SmRuAsO synthesis produced the cleanest sample (Figure \ref{fig:rietveld}).

\subsection{Physical properties}
\begin{figure}
\includegraphics[width=3.0in]{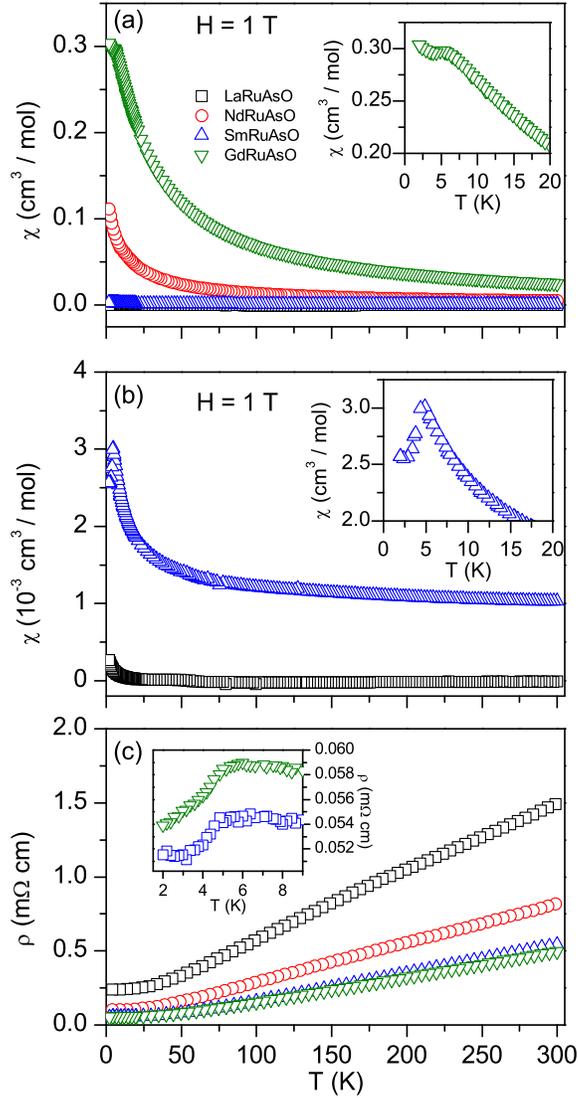}
\caption{\label{fig:mag-res}
(a,b) Temperature dependence of the magnetic susceptibility ($\chi\ = M/H$) per mole of formula units (note the different y-axis scales). The insets show the data for GdRuAsO (a) and SmRuAsO (b) near their Neel temperatures. (c) Temperature dependence of the resistivity ($\rho$) with the low temperature behavior for SmRuAsO and GdRuAsO in the inset.
}
\end{figure}

Results of magnetization and electrical resistivity measurements are shown in Figure \ref{fig:mag-res}. Magnetism in these compounds is dominated by the rare-earth element. The very small and nearly temperature independent susceptibility of LaRuAsO suggest that Ru does not have a local magnetic moment in these materials. The small upturn at low temperatures corresponds to an effective moment of 0.09 $\mu_B$ per formula unit, and is likely due to magnetic impurities.

The magnetic susceptibility ($\chi$) has a cusp at low temperatures for SmRuAsO and GdRuAsO, and displays Curie-Wiess (CW) behavior at higher temperatures. NdRuAsO shows CW behavior over the entire temperature range investigated. Effective moments ($\mu_{eff}$) and Weiss temperatures ($\theta$) determined from CW fits to the data from 50--300 K are listed in Table \ref{lattice-table}. The usual CW model is not sufficient for fitting the susceptibility data for SmRuAsO. This is not surprising since Sm$^{3+}$ is known to have closely space energy levels with different magnetic moments and strongly temperature dependent populations in the temperature ranges studied here. Effective moments in Table \ref{lattice-table} are consistent with the free ion values of 7.94 $\mu_B$ for Gd$^{3+}$ and 3.62 for Nd$^{3+}$, and similar to previous reports for the relevant Fe-compounds NdFeAsO (3.60 $\mu_B$ \cite{McGuire-LnFeAsO}) and GdFeAsO (7.83 $\mu_B$ \cite{Cui-GdFeAsO}). Weiss temperatures are negative, indicating antiferromagnetic interactions, and the cusps in $\chi$(T) near 5 K for SmRuAsO and GdRuAsO are attributed to antiferromagnetic ordering of rare earth moments. No evidence of magnetic ordering of Nd moments is observed above 2 K.

Electrical resistivity measurements (Figure \ref{fig:mag-res}c) indicate metallic behavior for these materials, similar to previous reports for LaRuAsO \cite{Chen-LnRuAsO}, CeFeAsO \cite{Chen-LnRuAsO}, SrRu$_2$As$_2$ \cite{Nath-AeRu2As2}, and BaRu$_2$As$_2$ \cite{Nath-AeRu2As2}. The magnitude of the resistivity ($\rho$) at room temperature decreases across the series from La to Gd. The inset in Figure \ref{fig:mag-res}c shows the effect of the magnetic ordering on the electrical resistivity in SmRuAsO and GdRuAsO. Reduced spin-disorder scattering below the transition is expected to be responsible for the decrease in $\rho$. Hall coefficients ($R_H$) measured at 2 K are listed in Table \ref{lattice-table}. The values are negative, indicating conduction dominated by electrons, and the magnitude decreases from $Ln$ = La to Gd. Carrier concentrations ($n_H$) inferred from these Hall coefficients from the simple one band formula $R_H$ = 1/$n_H$e range from $2\times10^{21}\ cm^{-3}$ for LaRuAsO to $1\times10^{22}\ cm^{-3}$ for GdRuAsO. The inferred carrier concentrations trend with the resistivity values near room temperature, and imply Hall mobilities of about 1--2 cm$^2$/Vs at room temperature.

\begin{figure}
\includegraphics[width=3.50in]{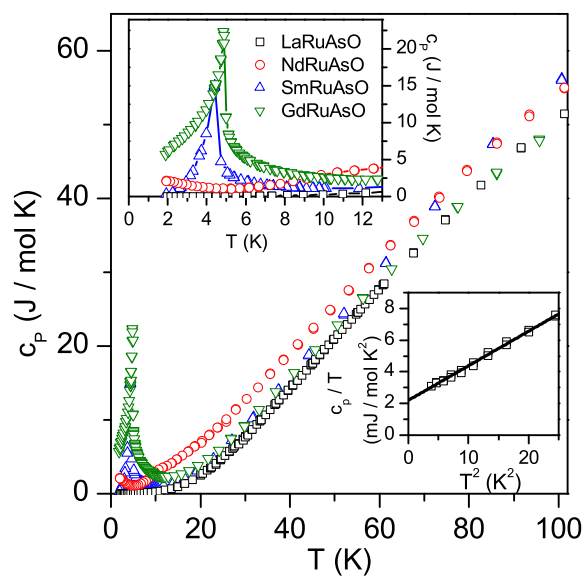}
\caption{\label{fig:hc}
Temperature dependence of the heat capacity per mole of formula unit below 100 K. The upper inset shows the sharp anomalies occurring in SmRuAsO and GdRuAsO at the magnetic ordering temperatures, and an upturn at the lowest temperatures for NdRuAsO suggesting ordering below 2 K. The lower inset shows $c_P/T$ vs. $T^2$ for LaRuAsO, and the linear fit used to determine the electronic heat capacity coefficient.
}
\end{figure}

To confirm the bulk nature of the magnetic phase transitions, heat capacity measurements were performed. The results are shown in Figure \ref{fig:hc}. Sharp anomalies are observed at the ordering temperatures for SmRuAsO and GdRuAsO. An upturn below about 4 K in the NdRuAsO data suggest the Nd magnetic moments may undergo a long range ordering transition near or below 2 K. NdFeAsO shows similar heat capacity behavior in this temperature range, with Nd moments ordering near 2.1 K \cite{Barker-NdFeAsO-hc}. A small anomaly is observed near 11 K in the NdRuAsO heat capacity data (Figure \ref{fig:hc}). Although this may be intrinsic, it is perhaps more likely due to the presence of a very small amount of NdAs, which has magnetic phase transition near 11 K with a large heat capacity anomaly \cite{NdAs-hc}. Based on the heat capacity data along with the resistivity and magnetic susceptibility results in Figure \ref{fig:mag-res} the antiferromagnetic ordering temperatures ($T_N$) are estimated to be 4.5 K for SmRuAsO and 5.0 K for GdRuAsO (Table \ref{lattice-table}). These are close to the rare-earth ordering temperatures in the related Fe materials: 4.1 K for GdFeAsO \cite{Wang-GdFeAsO} and 5--6 K in SmFeAsO \cite{Cimberle-SmFeAsO}.

The entropy change associated with the magnetic ordering in SmRuAsO and GdRuAsO can be determined by the integral $\int dTc_P^{mag}/T$. The magnetic heat capacity $c_P^{mag}$ is estimated by subtracting the heat capacity of LaRuAsO for the total heat capacity. Integrating up to T = 15 K gives an entropy per mole of lanthanide of 5.4 J/K$^2$/mol for SmRuAsO and 16.2 J/K$^2$/mol for GdRuAsO. The value obtained for GdRuAsO is close to R$ln(8)$ = 17.3 J/K$^2$/mol expected for $J = \frac{7}{2}$. The experimental value is likely underestimated due to the large value of $c_P$ remaining at the lowest temperature investigated here (Figure \ref{fig:hc}). For SmRuAsO, the entropy value is close to R$ln(2)$ =5.8 J/K$^2$/mol. This result is somewhat surprising, since $J = \frac{5}{2}$ is expected for the ground state of Sm. However, similar values have been obtained for SmFeAsO$_{1-x}$F$_x$ and attributed to crystal field splitting which results in an $J = \frac{1}{2}$ state at the lowest temperatures \cite{Barker-NdFeAsO-hc}.

For LaRuAsO, the electronic specific heat coefficient ($\gamma$) and the Debye temperature ($\theta_D$) can be determined from the plot of $c_P$/T vs. T$^2$ shown in the lower inset of Figure \ref{fig:hc}. Assuming $c_P = \gamma T + \beta T^3$ at low temperature, $\gamma$ = 2.2 mJ/K$^2$/mol-F.U. or 0.55 mJ/K$^2$/mol-atom is obtained. This is not a very high value, suggesting electron correlations are weak in these materials. The Debye temperature determined from $\beta$ is 330 K. These values can be compared to those reported for BaRu$_2$As$_2$ ($\gamma$ = 0.98 mJ/K$^2$/mol-atom, $\theta_D$ = 271 K) and SrRu$_2$As$_2$ ($\gamma$ = 0.82 mJ/K$^2$/mol-atom, $\theta_D$ = 271 K) \cite{Nath-AeRu2As2}.

\section{Summary}
The present study of the structural and physical properties of the Ru-based 1111 materials $Ln$RuAsO ($Ln$ = La, Nd, Sm, Gd) allows some interesting comparisons with isostructural and isoelectronic $Ln$FeAsO series. Structural differences include increased distortion of the transition metal coordination environment, and increased anisotropy in the response of the lattice to changing $Ln$ ionic radii. Both observations may be related to Ru-Ru repulsion in the $ab$-plane. Results of magnetization and heat capacity measurements indicate antiferromagnetic ordering of $Ln$ magnetic moments at 5.0 K in GdRuAsO, 4.5 K in SmRuAsO, and below 2 K in NdRuAsO, similar to values reported for Fe analogues. Evidence of the magnetic transitions in SmRuAsO and GdRuAsO is seen in the electrical resistivity. Together, resistivity and Hall effect measurements indicate metallic conduction dominated by electrons. No clear sign of Ru magnetism is observed in LaRuAsO, and analysis of low temperature heat capacity data suggest weak electron correlations in these materials. Estimates of entropy associated with the magnetic phase transitions give values close to R$ln(2)$ for SmRuAsO and R$ln(8)$ for GdFeAsO.

Research sponsored by the Materials Sciences and Engineering Division, Office of Basic Energy Sciences, U. S. Department of Energy.

\bibliographystyle{elsart-num}
%\bibliography{LnRuAsO}% Produces the bibliography via BibTeX.

\begin{thebibliography}{10}
\expandafter\ifx\csname url\endcsname\relax
  \def\url#1{\texttt{#1}}\fi
\expandafter\ifx\csname urlprefix\endcsname\relax\def\urlprefix{URL }\fi

\bibitem{Quebe}
P.~Quebe, L.~J. Terb{\"{u}}e, W.~Jeitschko, J. Alloys Compd. 302 (2000) 70.

\bibitem{Kamihara}
Y.~Kamihara, T.~Watanabe, M.~Hirano, H.~Hosono, J. Am. Chem. Soc. 130 (2008)
  3296.

\bibitem{Johnston-review}
D.~C. Johnston, Advances in Physics 59 (2010) 803.

\bibitem{Sefat-review}
A.~S. Sefat, D.~J. Singh, MRS Bulletin 36 (2011) 614.

\bibitem{Schnelle-SrFeRuAs}
W.~Schnelle, A.~Leithe-Jasper, R.~Gumeniuk, U.~Burkhardt, D.~Kasinathan,
  H.~Rosner, Phys. Rev. B 79 (2009) 214516.

\bibitem{Sharma-BaFeRuAs}
S.~Sharma, A.~Bharathi, S.~Chandra, V.~{Raghavendra Reddy}, S.~Paulraj, A.~T.
  Satya, V.~S. Sastry, A.~Gupta, C.~S. Sundar, Phys. Rev. B 81 (2010) 174512.

\bibitem{McGuire-PrFeRuAsO}
M.~A. McGuire, D.~J. Singh, A.~S. Sefat, B.~C. Sales, D.~Mandrus, J. Solid
  State Chem. 182 (2009) 2326.

\bibitem{Pallecchi-LaFeRuAsO}
I.~Pallecchi, F.~Bernardini, M.~Tropeano, A.~Palenzona, A.~Martinelli,
  C.~Ferdeghini, M.~Vignolo, S.~Massidda, M.~Putti, Phys. Rev. B 84 (2011)
  134524.

\bibitem{Pal-SmGdFeRuAsO}
A.~Pal, A.~Vajpayee, V.~P.~S. Awana, M.~Husain, H.~Krishan, Physica C 470
  (2010) S491.

\bibitem{Tropeano-SmFeRuAsOF}
M.~Tropeano, M.~R.~C. adn C.~Ferdeghini, G.~Lamura, A.~Martinelli,
  A.~Palenzona, I.~Pallecchi, A.~Sala, I.~Sheikin, F.~Bernardini, M.~Monni,
  S.~Sassidda, M.~Putti, Phys. Rev. B 81 (2010) 184504.

\bibitem{Jeitschko-LaRu2P2}
W.~Jetischko, R.~Glaum, L.~Boonk, J. Solid State Chem. 69 (1987) 93.

\bibitem{Nath-AeRu2As2}
R.~Nath, Y.~Singh, D.~C. Johnston, Phys. Rev. B 79 (2009) 174513.

\bibitem{Chen-LnRuAsO}
G.-F. Chen, Z.~Li, D.~Wu, J.~Dong, G.~Li, W.-Z. Hu, P.~Zheng, J.-L. Luo, N.-L.
  Wang, Chin. Phys. Lett. 25 (2008) 2235.

\bibitem{Fullprof}
J.~{Rodriguez-Carvajal}, Physica B 192 (1993) 55, program available at
  www.ill.fr/dif/Soft/fp/.

\bibitem{Zhao-Ce}
J.~Zhao, Q.~Huang, C.~{de la Cruz}, S.~Li, J.~W. Lynn, Y.~Chen, M.~A. Green,
  G.~F. Chen, G.~Li, Z.~Li, J.~L. Luo, N.~L. Wang, P.~Dai, Nature Materials 7
  (2008) 953.

\bibitem{Kreyssig}
A.~Kreyssig, M.~A. Green, Y.~Lee, G.~D. Samolyuk, P.~Zajdel, J.~W. Lynn, S.~L.
  Bud'ko, M.~S. Torikachvili, N.~Ni, S.~Nandi, J.~B. Le{\~{a}}o, S.~J. Poulton,
  D.~N. Argyriou, B.~N. Harmon, R.~J. McQueeney, P.~C. Canfield, A.~I. Goldman,
  Phys. Rev. B 78 (2008) 184517.

\bibitem{Nitsche-LnFeAsO}
F.~Nitsche, A.~Jesche, E.~Hieckmann, T.~Doert, M.~Ruck, Phys. Rev. B 82 (2010)
  134514.

\bibitem{Shannon}
R.~D. Shannon, Acta Crystallographica Section A 32 (1976) 751.

\bibitem{McGuire-LnFeAsO}
M.~A. McGuire, R.~P. Hermann, A.~S. Sefat, B.~C. Sales, R.~Jin, D.~Mandrus,
  F.~Grandjean, G.~J. Long, New J. Phys. 11 (2009) 025011.

\bibitem{Cui-GdFeAsO}
Y.~Cui, Y.~Chen, C.~Cheng, Y.~Yang, Y.~Zhang, Y.~Zhao, J. Supercond. Novel Mag.
  23 (2010) 625.

\bibitem{Barker-NdFeAsO-hc}
P.~J. Barker, S.~R. Giblin, F.~L. Pratt, R.~H. Liu, G.~Wu, X.~H. Chen, M.~J.
  Pitcher, D.~R. Parker, S.~J. Clarke, S.~J. Blundell, New J. Phys. 11 (2009)
  025010.

\bibitem{NdAs-hc}
A.~Aeby, F.~Hulliger, B.~Natterer, Solid State Comm. 13 (1973) 1365.

\bibitem{Wang-GdFeAsO}
P.~Wang, Z.~M. Stadnik, C.~Wang, G.-H. Cao, Z.-A. Xu, J. Phys.: Condens. Matter
  22 (2010) 145701.

\bibitem{Cimberle-SmFeAsO}
M.~R. Cimberle, F.~Canepa, M.~Ferritti, A.~Martinelli, A.~Palenzona, A.~S.
  Siri, C.~Tarantini, M.~Tropeano, C.~Ferdeghini, J. Magn. Magn. Mater. 321
  (2009) 3024.

\end{thebibliography}

\end{document}